\documentclass[runningheads,a4paper]{llncs}
\usepackage{amssymb}
\usepackage{graphicx}
\begin{document}

\mainmatter
\title{Towards protein-protein docking with significant structural changes using CABS-dock}
\titlerunning{Towards protein-protein docking using CABS-dock}
\author{Maciej Pawe\l{} Ciemny, Mateusz Kurcinski, Andrzej Kolinski and Sebastian Kmiecik}
\authorrunning{Towards protein-protein docking using CABS-dock}
\institute{University of Warsaw, Chemistry Department,\\ Pasteura 1, 02-093 Warsaw,\\ Poland
}
\maketitle

\begin{abstract}
The protein-protein interactions (PPIs) are crucial for understanding the majority of cellular processes. PPIs play important role in gene transcription regulation, cellular signaling, molecular basis of immune response and more. Moreover, a disruption of these mechanisms is frequently postulated as a possible cause of diseases such as Alzheimer’s or cancer. For many of biologically relevant cases the structure of protein-protein complexes remain unknown. Therefore computational techniques, including molecular docking, have become a valuable part of drug discovery pipelines. Unfortunately, none of the widely used protein-protein docking tools is free from serious limitations. Typically, in docking simulations the protein flexibility is either completely neglected or very limited. Additionally, some knowledge of the approximate location and/or the shape of the active site is also required. Such limitations arise mostly from the enormous number of degrees of freedom of protein-protein systems. In this paper, an efficient computational method for protein-protein docking is proposed and initially tested on a single docking case. The proposed method is based on a two-step procedure. In the first step, CABS-dock web server for protein-peptide docking is used to dock a peptide, which is the appropriate protein fragment responsible for the protein-protein interaction, to the other protein partner. During peptide docking, no knowledge about the binding site, nor the peptide structure, is used and the peptide is allowed to be fully flexible. In the second step, the docked peptide is used in the structural adjustment of protein complex partners. The proposed method allowed us to obtain a high accuracy model, therefore it provides a promising framework for further advances.
\end{abstract}

\section{Introduction}
As recent studies show, a vast majority of important cellular functions are regulated and mediated by protein-protein interactions (PPIs). It has also been demonstrated that the peptide-protein interactions are the key-mediators of PPIs [1]. Those include both interactions of a protein receptor with small, flexible peptides as well as with linear motifs/segments that belong to longer protein chains. This fact leads to a hypothesis that the protein-protein complex structure may be (at least -- to some extent) determined by interaction with properly chosen binding motifs from its complex partner [2]. Those findings encourage a new approach to protein-protein docking that uses information on interactions between protein and linear segments extracted from its complex partner to reconstruct the whole complex. Here we present a simple modeling scheme that combines this newly formulated ideas with a well-tested method for flexible protein-peptide docking -- CABS-dock [3, 4].

\section{Description of the Method}
The presented approach may be divided into three main modeling steps: 
\begin{itemize}
\item Reduction of the protein-protein docking problem to protein-peptide docking. This starts from arbitrary selection of the receptor protein and bounded (more likely the smaller one) protein, followed by the selection of ‘hot segment(s)’ -- protein fragment(s) responsible for the interaction between proteins [5, 6]. 
\item Protein-peptide docking of ‘hot segment(s)’ i.e. peptides using CABS-dock
\item Reconstruction and adjustment of the remaining receptor structure to the docked peptide-like fragment.
\end{itemize}

Each of the steps that compose the protocol is a demanding challenge on its own. The first issue is the method to choose the peptide -- linear motif --  from the interaction site of the complex. The final method should only use the information about non-bound structures (experimental or predicted) of the proteins taking part in the complex formation. The second step is mainly addressed by the CABS-dock method. CABS-dock does not require any information on the docking site on the receptor protein, which makes it perfectly suited for the proposed procedure. For details on the CABS-dock procedure and its performance, see [3, 4]. The major challenge in this step lies in the accurate scoring of the generated protein-peptide complexes. Finally, the reconstruction of the complex structure, based on short segments derived from protein-peptide docking, should be able to take into account significant changes of the protein partners in respect to the unbound structures.

In the next section, we present a simple realization of this procedure together with an example of reconstructed complex.

\section{Preliminary Results}
Here we present a short case-study of a successful complex structure reconstruction performed using the presented scheme. The example was chosen from the non-redundant docking Benchmark 3.032 [7] where it was assigned to the “medium difficulty” group (relevant PDB codes: cbound complex -- 1ACB, non-bound proteins: 2CGA -- Chymotrypsin,  1EGL -- Eglin C).

\subsection{Step 1. Selection of 'Hot Segment' (Peptide)}
To satisfy the need for choosing a linear segment of the “binding” protein we used a ten amino-acids long linear motif (decamer) determined by N. London et al. [5] for this complex. The “hot segments”, that mostly contribute to the protein-protein interaction energy of the complex were determined using the Rosetta method for interaction energy calculation. For exact localization of the decamer in the complex structure, see Figure 1.

\begin{figure}
\centering
\includegraphics[height=6.2cm]{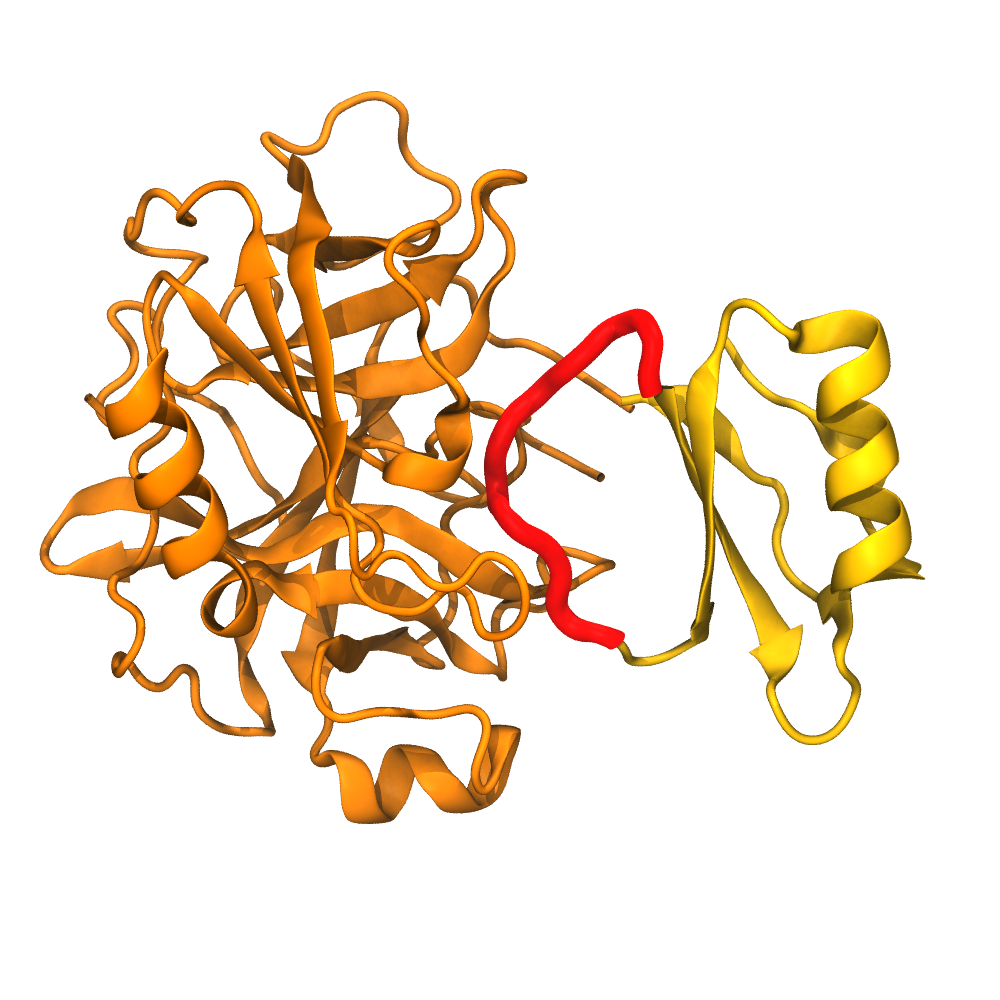}
\caption{The decamer (red) chosen as a linear motif of eglin C (yellow ribbon ) for docking to the receptor -- chymotrypsin (orange ribbon).}
\label{rys0}
\end{figure}

\subsection{Step 2. Protein-peptide docking using CABS-dock}
The CABS-dock server (available at http://biocomp.chem.uw.edu.pl/CABSdock) was used to generate a ranked set of peptide-protein complex models. Figure 2 presents the resulting structures together with the receptor protein. Please note that the method does not use any knowledge on the location of the binding site of the receptor. For further modeling steps, we selected the most accurate model from 10 top-ranked models -- characterized by the RMSD value of 5.66\AA~(as calculated to the respective element of the protein in stable complex).

\begin{figure}
\centering
\includegraphics[height=6.2cm]{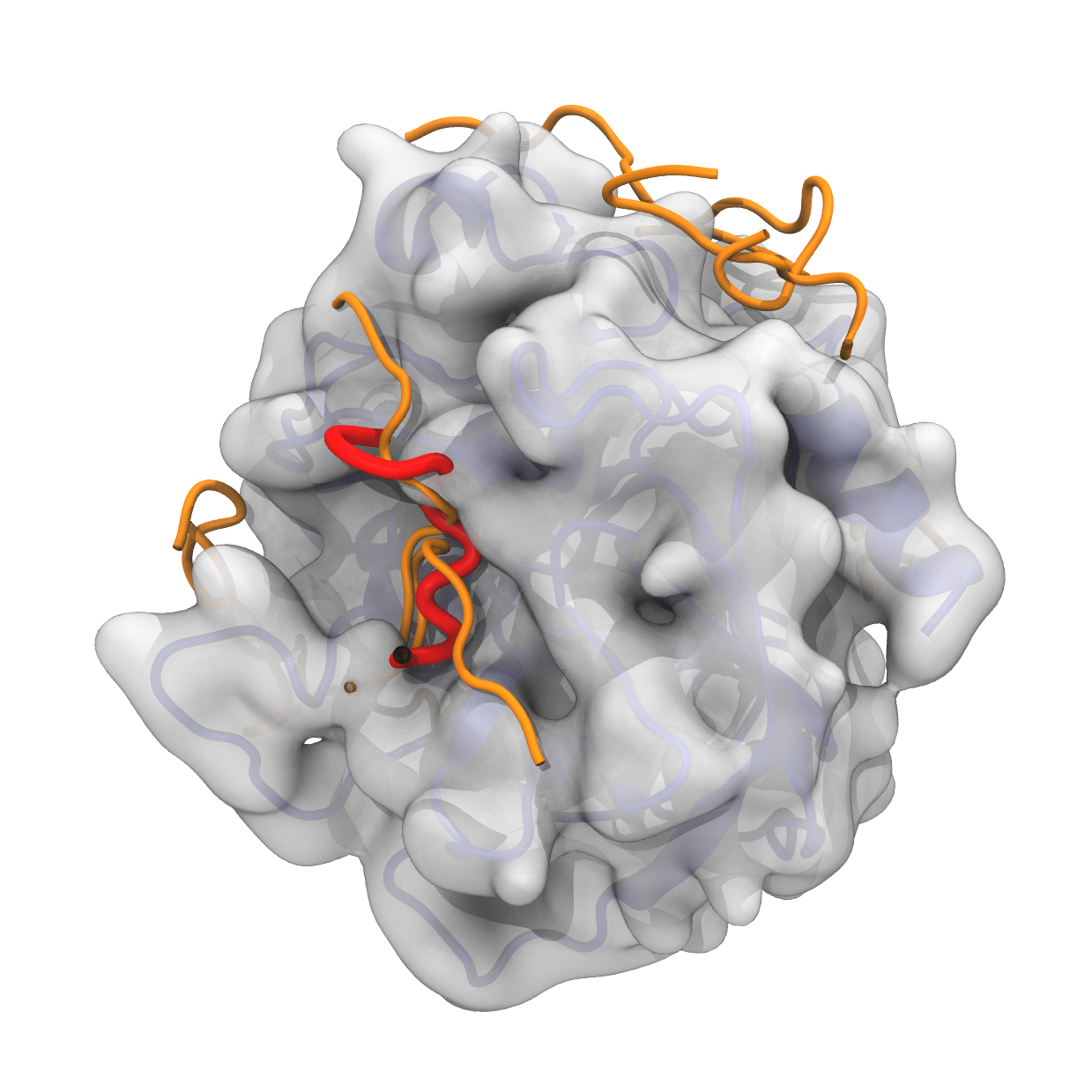}
\caption{The ten top-ranked models resulting from protein-peptide docking of the chosen linear motif. The peptides are shown in orange, the protein receptor (chymotrypsin) in white with blue ribbon representation of elements of the secondary structure. The most accurate model (RMSD 5.66\AA) of eglin C linear motif is marked in red.}
\label{rys1}
\end{figure}

\subsection{Step 3. Adjustment of protein-protein complex based on predicted protein-peptide complex}
The adjustment of protein-protein complex is realized in two stages. Firstly, eglin C is docked to chymotrypsin based on the superimposition of eglin C hot segment with the peptide obtained in the CABS-dock docking. The resulting complex structure of RMSD 2.87\AA~is presented in Figure 3.

\begin{figure}
\centering
\includegraphics[height=6.2cm]{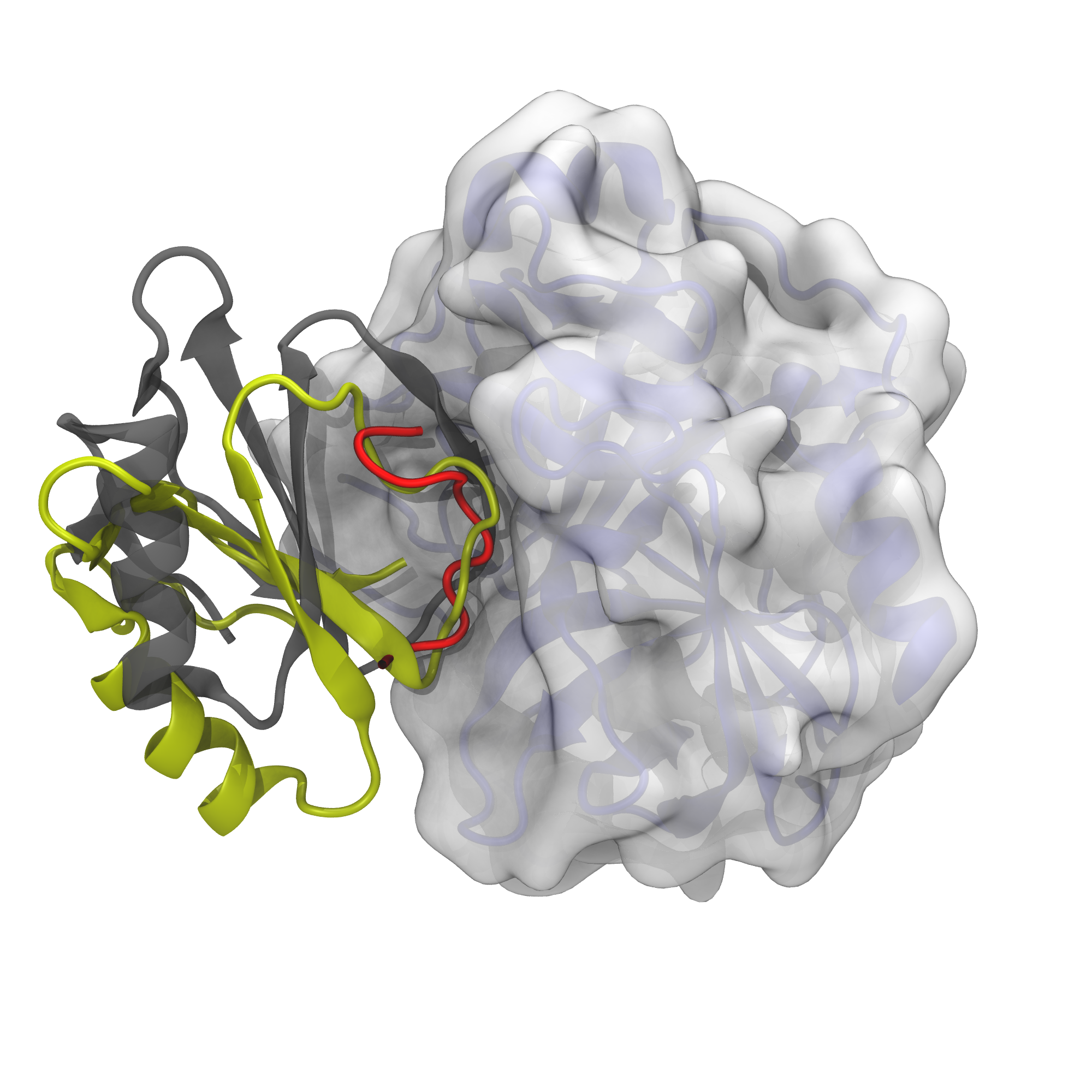}
\caption{The protein complex reconstructed by simple superimposition of the interacting protein on the docked peptide (RMSD 5.66\AA, shown in red in the illustration). The protein receptor -- chymotrypsin - is shown in white with blue ribbon representation of elements of the secondary structure, eglin C - in green. The crystallographic structure of the complex is given for reference in dark grey transparent representation.}
\label{rys2}
\end{figure}

The obtained protein-protein complex structure was further refined using CABS modeling scheme. What is important in this context CABS model is a well-tested tool for efficient modeling of large-scale conformational transitions (including protein folding [8-10], protein flexibility [11-13], peptide binding [14-18] and modeling of protein complexes [19-21]). In the set of clusters of resulting structures we have found one characterized by average RMSD of 1.75 \AA. The comparison between the crystallographic structure of the complex and the CABS-based prediction is presented in the Figure 4.

\begin{figure}
\centering
\includegraphics[height=6.2cm]{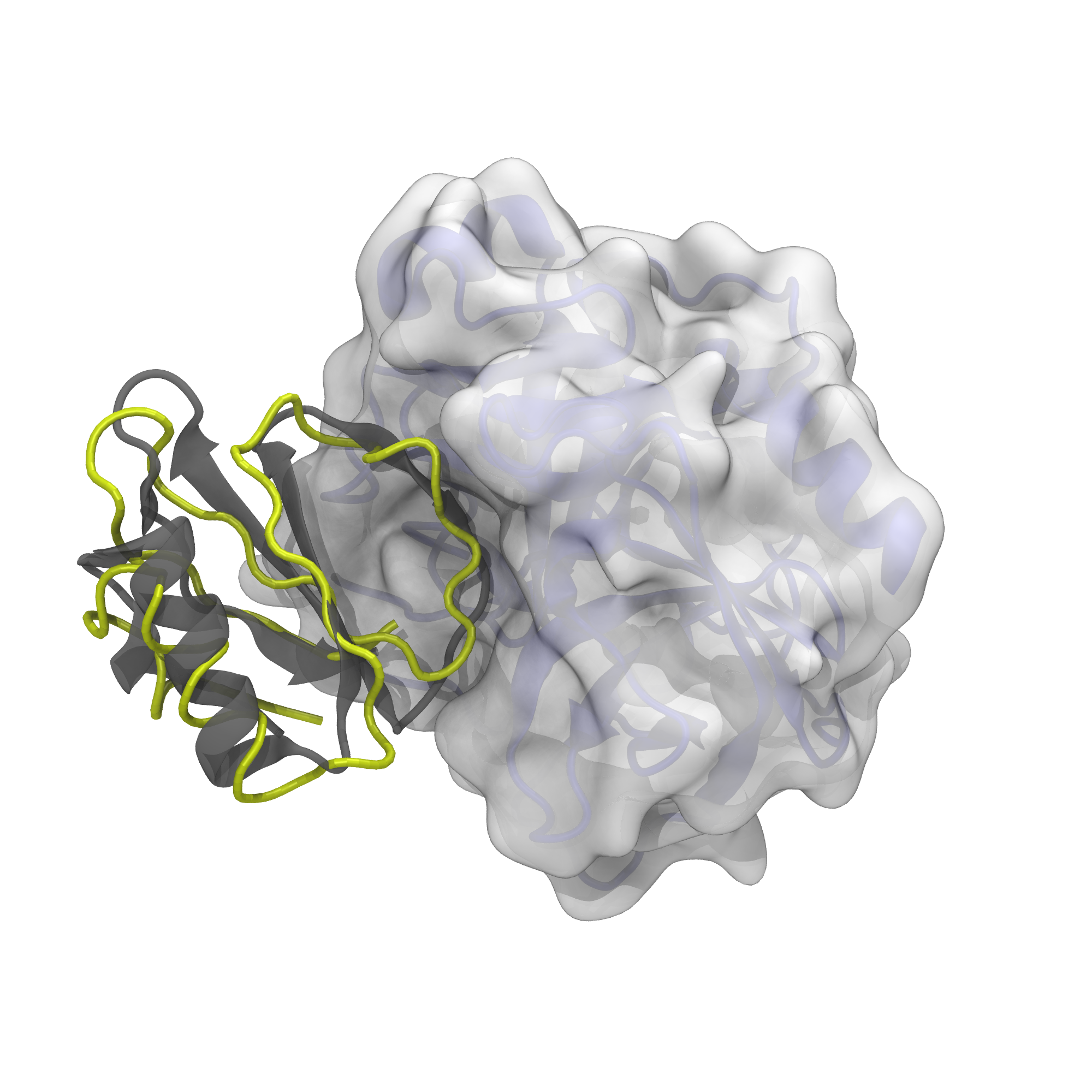}
\caption{The refined protein-protein complex (RMSD 1.75\AA). Chymotrypsin is shown in white with blue ribbon representation of elements of the secondary structure, eglin C - in green. The crystallographic structure of the complex is given for reference in dark grey transparent representation.}
\label{rys3}
\end{figure}

\section{Future Perspectives}
In this work, we present an example how the protein-protein complex structure may be predicted based on the results of protein-peptide docking using CABS-dock. Our CABS-dock-based method yielded a model with RMSD value 1.75\AA~--which is a good starting point for a more precise modeling. Combining our method with a high-resolution refinement and/or scoring will possibly lead to greater accuracy of the final model [22, 23]. The results of this initial validation test are very promising, however, we expect that further development of the presented methodology will be a challenging task. First of all, the scoring of protein-peptide complexes is a difficult problem and requires tools dedicated solely to this purpose. Similarly, the last modeling phase -- the adjustment of protein-protein complex -- would also require improvements to achieve the best possible scoring. The presented protein-protein docking scheme can be extended to incorporate alternative pathways that rely on methods other than CABS-dock. The resulting structures may be then compared and cross-validated to increase the credibility of obtained models.

\section{Acknowledgments}
The authors acknowledge support from the National Science Center grant [MAESTRO 2014/14/A/ST6/00088].

\end{document}